\documentclass[sigconf]{acmart}
\usepackage{array}
\usepackage{units}
\usepackage{graphicx}
\usepackage{subfig}
\newcolumntype{C}[1]{>{\centering\arraybackslash}p{#1}}

\AtBeginDocument{%
  \providecommand\BibTeX{{%
    \normalfont B\kern-0.5em{\scshape i\kern-0.25em b}\kern-0.8em\TeX}}}

\acmYear{2021}\copyrightyear{2021}
\setcopyright{rightsretained}
\acmConference[DICPS '21]{The Workshop on Data-Driven and Intelligent Cyber-Physical Systems}{May 18, 2021}{Nashville, TN, USA}
\acmBooktitle{The Workshop on Data-Driven and Intelligent Cyber-Physical Systems (DICPS '21), May 18, 2021, Nashville, TN, USA}
\acmPrice{}
\acmDOI{10.1145/3459609.3460530}
\acmISBN{978-1-4503-8445-2/21/05}

\begin{document}

\title[Integrated Framework of Dynamics, Instabilities, Energy Models, and Sparse Flow Controllers]{Integrated Framework of Vehicle Dynamics, Instabilities, Energy Models, and Sparse Flow Smoothing Controllers}

\settopmatter{printacmref=false}

\author{
    Jonathan W. Lee\textsuperscript{\textdagger}, 
    George Gunter\textsuperscript{\textdaggerdbl}, 
    Rabie Ramadan\textsuperscript{\textasteriskcentered}, 
    Sulaiman Almatrudi\textsuperscript{\textdagger}, 
    Paige Arnold\textsuperscript{\textsection}, 
    John Aquino\textsuperscript{\textdagger},
    William Barbour\textsuperscript{\textdaggerdbl}, 
    Rahul Bhadani\textsuperscript{\textbardbl}, 
    Joy Carpio\textsuperscript{\textdagger}, 
    Fang-Chieh Chou\textsuperscript{\textdagger}, 
    Marsalis Gibson\textsuperscript{\textdagger}, 
    Xiaoqian Gong\textsuperscript{\textsection}, 
    Amaury Hayat\textsuperscript{\textsection}, 
    Nour Khoudari\textsuperscript{\textasteriskcentered}, 
    Abdul Rahman Kreidieh\textsuperscript{\textdagger},
    Maya Kumar\textsuperscript{\textdaggerdbl}, 
    Nathan Lichtl\'e\textsuperscript{\P},
    Sean McQuade\textsuperscript{\textsection}, 
    Brian Nguyen\textsuperscript{\textdagger}, 
    Megan Ross\textsuperscript{\textasteriskcentered}, 
    Sydney Truong\textsuperscript{\textsection}, 
    Eugene Vinitsky\textsuperscript{\textdagger}, 
    Yibo Zhao\textsuperscript{\textdagger}, 
    Jonathan Sprinkle\textsuperscript{\textbardbl}, 
    Benedetto Piccoli\textsuperscript{\textsection}, 
    Alexandre M. Bayen\textsuperscript{\textdagger}, 
    Daniel B. Work\textsuperscript{\textdaggerdbl}, 
    Benjamin Seibold}
\authornote{
    --Temple University\\
    \textsuperscript{\textdagger}--University of California, Berkeley\\
    \textsuperscript{\textdaggerdbl}--Vanderbilt University\\
    \textsuperscript{\textsection}--Rutgers University-Camden\\
    \textsuperscript{\textbardbl}--University of Arizona\\
    \textsuperscript{\P}--\'Ecole normale sup\'erieure Paris-Saclay, Paris-Saclay University}
\email{circles.consortium@gmail.com}

\renewcommand{\shortauthors}{Lee, \emph{et al.}}

\begin{abstract}
This work presents an integrated framework of: vehicle dynamics models, with a particular attention to instabilities and traffic waves; vehicle energy models, with particular attention to accurate energy values for strongly unsteady driving profiles; and sparse Lagrangian controls via automated vehicles, with a focus on controls that can be executed via existing technology such as adaptive cruise control systems. This framework serves as a key building block in developing control strategies for human-in-the-loop traffic flow smoothing on real highways. In this contribution, we outline the fundamental merits of integrating vehicle dynamics and energy modeling into a single framework, and we demonstrate the energy impact of sparse flow smoothing controllers via simulation results.
\end{abstract}



\begin{teaserfigure}
  \includegraphics[width=\textwidth]{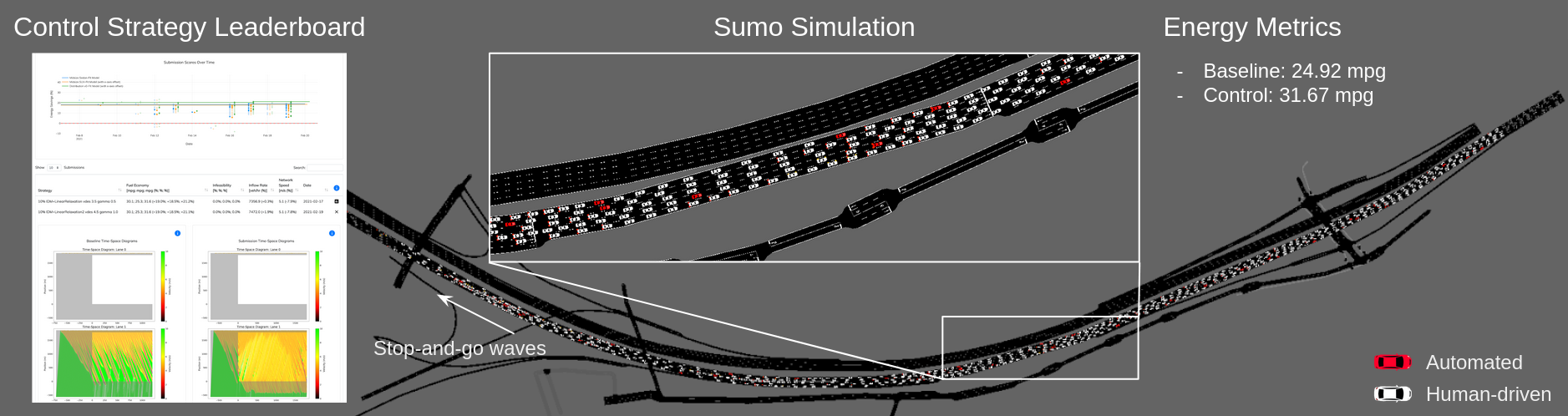}
  \caption{Traffic waves generated by human driving increase the energy consumption of traffic flow. A small fraction of well-controlled automated vehicles can smooth the flow and the reduce energy consumption.}
  \Description{traffic waves in a microsimulation.}
  \label{fig:teaser}
\end{teaserfigure}

\maketitle

\section{Introduction}
An emerging research direction is the development of concepts and technologies that enable energy optimization of traffic flow using automated vehicle technologies to control the overall traffic, by exploiting the influence of these vehicles on the bulk traffic flow. In this endeavour it is crucial to correctly model system-level impacts when sparse controllers (\emph{i.e.}, automated vehicles) close the loop in the real-world application. Moreover, because most of the agents remain human-controlled vehicles, the human-in-the-loop aspects (particularly the reactions of humans to the controllers) are critical for successful control design.

Because such automated vehicles may have different objectives compared to human drivers or classical \emph{Adaptive Cruise Controlled} (ACC) vehicles, a critical enabler of real-world deployment is an integrative simulation/software tool that combines an accurate capture of vehicle dynamics, vehicle energy modeling, as well as controllers, in a way that both micro-scale and macro-scale flow features are reproduced with suitable fidelity. Such tools are needed but not readily available to support research and development.

To illustrate the importance of human driving behavior on traffic flow, the seminal work \cite{Sugiyamaetal2008} showed experimentally that human-controlled traffic can exhibit ``phantom jams'' in which a traffic jam emerges, not from an outside influence such as an accident or lane-reduction, but from the collective behavior of drivers. 
Phantom jams are an important area of traffic control research because they represent a true inefficiency in the flow in which both system throughput is degraded and average fuel efficiency declines~\cite{stern2017dissipation,wu2019tracking}, and frequently arise in non-automated (human) traffic flows~\cite{Sugiyamaetal2008}.
It was subsequently shown \cite{stern2017dissipation} that these jams could be effectively ``smoothed out'' with a single automated vehicle on a ring of 20+ other human drivers. The result was that throughput, and notably, energy efficiency of the system were both improved.

Since then, a number of theoretical works have expanded on techniques for using sparsely adopted \emph{connected and automated vehicles} (CAVs) to smooth traffic \cite{delle2019feedback,wu2018stabilizing}. Despite the potential for CAVs to significantly improve operation conditions for congested traffic flow, currently available vehicles in the form of ACC do not seem to display such a tendency. Across \cite{milanes2014cooperative,gunter2019model,gunter2020commercially}, several commercially available ACCs were tested experimentally, and all such systems actually contributed to phantom jams. As a result, there is a gap between where traffic flow theory and experimentation suggest CAVs could be and where they are in practice.

In order to enable addressing this disconnect, this work develops a high fidelity simulation environment that allows users to quickly implement and test CAV control schemes in complicated traffic environments to assess their capacity to smooth waves and improve energy consumption. A number of microsimulation packages exist, both proprietary and open-source. We build on the widely used open source traffic simulation environment \emph{Simulation of Urban MObility} (SUMO)~\cite{SUMO}, but recognize it does not directly produce stop-and-go waves (a core feature of traffic that leads to wasted energy). Additionally, we also recognize that suitable models to estimate energy from these trajectories are needed. On one hand, aggregated models such as the US EPA MOVES~\cite{MOVES} model produce energy estimates, but based on average speeds on road segments. On the other extreme, models such as the US DOE Autonomie~\cite{AUTONOMIE} are highly detailed but computationally costly and not open-source compatible.
Recognizing the gap between the tools, here we outline our preliminary work to producing a set of software tools that enables researchers concerned with the development of vehicle controllers to improve the energy efficiency of traffic to develop, test, and benchmark their ideas. The resulting tools take care of the intricacies of non-equilibrium traffic flow theory (\emph{e.g.}, producing phantom jams) while also appropriately approximating vehicle energy dynamics compatible with the traffic dynamics. We demonstrate this energy modeling on generic vehicle types and a Toyota RAV4 vehicle, which is the vehicle platform we intend to use later for large scale traffic control. We also incorporate clear metrics for benchmarking through a leaderboard. The leaderboard provides common traffic scenarios and reports metrics (\emph{e.g.}, system-level energy and flow metrics) to assess controllers' effect on the flow dynamics on a system-level scale. While the tools presented here are in a preliminary stage, they are already being used by our multi-institution consortium that includes more than 50 researchers building out these tools~\cite{CIRCLES_webpage}.

The remainder of the article is as follows. In Section \ref{sec:waves} we discuss the challenges to integrate stop-and-go traffic dynamics into SUMO. In Section \ref{sec:energy} we outline our work to build simplified energy models that approximate Autonomie models and thus are suitable for integration with SUMO. In Section \ref{sec:integration} we discuss the integration efforts and a proof-of-concept case study. Conclusions and future perspectives are given in Section \ref{sec:conclusions}.

\section{Traffic Dynamics Modeling}
\label{sec:waves}
Traffic is primarily simulated and understood via two different paradigms: \textit{macroscopic} and \textit{microscopic}. In macroscopic modeling, aggregate states of traffic flow are modeled over space and time. Relevant quantities are the \textit{flow rate}, \textit{speed}, and \textit{density} of traffic, and typically partial differential equations are used for simulation, resembling fluid dynamics.

In contrast, the microscopic view models interactions between individual vehicles. These techniques use the \textit{speed}, \textit{spacing-gap}, and \textit{speed-difference} of each vehicle in the network to advance the simulation as a system of ordinary differential equations. Since the primary focus of this work is to understand how controllers implemented on a few CAVs can improve traffic flow, the microscopic paradigm is a natural choice. It allows for high fidelity modeling of individual vehicles, both from a control design perspective and from an energy consumption perspective. 

Of particular interest for using automated vehicles to smooth traffic flow is the car-following behavior. Typically, \emph{car-following models} (CFMs) are either first-order or second-order ODEs, corresponding to either modeling the speed of a vehicle directly or prescribing its acceleration. Suitable for traffic waves are second-order models of the form
\begin{equation}\label{eq:generic_cfm}
\begin{array}{rl}
    & \dot{v}(t) = f_\text{CFM}(s(t), v(t), \Delta v(t))\\
    & \dot{s}(t) = \Delta v(t)
\end{array}
\end{equation}
where $\dot{v}(t)$ is the acceleration of an ego vehicle, governed by the equation $f_\text{CFM}$, $s(t)$ is the space-gap to the lead vehicle, $v(t)$ is the vehicle's speed, and $\Delta v(t)$ is the inter-vehicle speed-difference. Generally, CFMs are used to describe the driving behavior of human drivers, with some popular choices being the Gipps Model~\cite{gipps1981behavioural}, the Optimal-Velocity model~\cite{nakayama2016quantitative}, and the Intelligent Driver Model (IDM)~\cite{treiber2000congested}. In this work we use the IDM for human driving, which is of the form
\begin{equation}\label{eq:IDM}
\begin{array}{rl}
    & f_\text{IDM}(s(t), v(t), \Delta v(t)) = a\left[1-\left(\frac{v}{v_{0}}\right)^{\delta}\!\!-\left(\frac{s^{*}\left(v,\Delta v\right)}{s}\right)^{2}\right]\\
    & s^{*}(v,\Delta v) = s_{0}+vT+\frac{\max\{0,v\Delta v\}}{2\sqrt{ab}}\;,\\
\end{array}
\end{equation}
and $a$, $b$, $v_{0}$, $\delta$, $T$, and $s_{0}$ are model parameters that influence driver behavior. Moreover, the acceleration \eqref{eq:IDM} is replaced by zero whenever $f_\text{IDM}<0$ and $v=0$. 

Of central interest in this work are phantom jams, in which small disturbances to an otherwise steady traffic flow can translate into large slow-downs that move move backwards in the flow. Mathematically, phantom jams can be interpreted as dynamic instabilities that grow into nonlinear waves~\cite{FlynnKasimovNaveRosalesSeibold2009}. In suitable scenarios, the (in)stability of a uniform flow is equivalent to the concept of \emph{string-stability} \cite{wilson2011car}. To determine whether a CFM can exhibit phantom jams due to string-instability, a linear stability analysis using partial derivatives can be applied to derive a transfer function:
\begin{equation*}
    \ddot{y}_{i}(t) = \alpha_1 (y_{i-1} - y_i) - \alpha_2 \dot{y}_{i} + \alpha_3 \dot{y}_{i-1}\;,\\
\end{equation*}
\begin{equation*}
     \text{where~~}
     \alpha_1 = \tfrac{\partial f}{\partial s}\,,\
     \alpha_2 = \tfrac{\partial f}{\partial (\Delta v)} - \tfrac{\partial f}{\partial v}\,,\
     \alpha_3 = \tfrac{\partial f}{\partial (\Delta v)}\;,\\
\end{equation*}
\begin{equation*}
    \text{and~~} F(\omega) = \frac{\alpha_1 +\alpha_3 \omega}{\alpha_1 +\alpha_2\omega + \omega^2}\;.\\
\end{equation*}
Here $F(\omega)$ represents the amplification of a perturbation of frequency $\omega$, and $y(t)$ is a linearization of the CFM. The system is string stable if $|F(\omega)|\leq 1 \; \forall \omega \in i\mathbb{R}$, which is equivalent to the simple algebraic criterion
\begin{equation}
    \label{eq:stability_condition}
    \lambda \geq0 \quad\text{where}\quad
    \lambda = \alpha_2^2 - \alpha_3^2 - 2\alpha_1\;.
\end{equation}
Traffic modeled with a given CFM is then able to exhibit phantom jams when \eqref{eq:stability_condition} is violated (see \cite{cui2017stabilizing} for more details).


Leveraging this stability theory, a microsimulation environment is set up in SUMO based on the IDM \eqref{eq:IDM} to model traffic streams of human drivers which exhibit phantom jams, with the goal of designing CAV controllers to then smooth those waves. To do this, IDM parameters are chosen such that (i)~a realistic fundamental diagram is reproduced; and (ii)~at/near the critical density, a transition from stability to instability occurs with growth rates that grow waves from small perturbations on a time scale of 10s--30s. We choose
$a=1.3$ m/s\textsuperscript{2},
$b=2$ m/s\textsuperscript{2},
$v_{0}=30$ m/s,
$\delta=4$,
$T=1\textrm{s}$, and
$s_{0}=1\textrm{m}$.


The integrator time step for the simulation is chosen small enough so that the stability criterion \eqref{eq:stability_condition} and wave behavior from the IDM are well-captured, but that it still allows for efficient simulation. Moreover, strict adherence to speed/acceleration/gap bounds of the discrete time-stepping system is guaranteed via standard fail-safes. Random noise is added to the prescribed IDM acceleration values to trigger any dynamic instabilities (the noise effects are properly filtered out when feeding accelerations into the energy models below). Finally, lane changing, geometries, vehicle inflow spawning, and routing are all handled using SUMO's existing capabilities, while the aforementioned custom car-following logic is implemented using the software package Flow~\cite{flow}, and further implementation details are described in Section \ref{sec:integration}. This framework leverages existing modeling work of the microsimulation to handle ``real'' traffic, while at the same time allowing for high fidelity and principled choice/development of a CFM that satisfies the critical requirement that it produces systematic instabilities and waves that resemble real stop-and-go traffic.

\section{Energy Models}
\label{sec:energy}
This project requires vehicle energy models that are (a)~accurate for highly non-constant velocities; (b)~representative of a variety of vehicle types; yet also (c)~structurally simple enough to allow for fast evaluation, use in optimization/control, and open source implementation. To meet requirements (a) and (b), the Autonomie software \cite{ANL} is employed on a selection of representative vehicles (Table~\ref{table:model_list_and_errors}), and a systematic model-reduction procedure is devised to generate simple fitted models that meet requirement (c).

\textbf{Vehicle portfolio:}
The selected vehicles are chosen to satisfy diversity and prevalence in the US. The top 5 vehicles in Table~\ref{table:model_list_and_errors} represent generic vehicles that are representative for current market vehicles of that specific type \cite{ANL}. Toyota RAV4 is chosen because this is the consortium's \cite{CIRCLES_webpage} intended primary controller vehicle used for flow smoothing. Each vehicle model represents a class of vehicles that have comparable weight (with load assumed half full) and fuel consumption characteristics. The road share in Table~\ref{table:model_list_and_errors} is obtained as follows: (i)~share of trucks (Class3 PND) vs.\ passenger vehicles (rest) from TN DoR \cite{TDOR} vehicle registration data (as future field tests are expected in TN); (ii)~distribution within passenger vehicles from CNCDA \cite{CNCDA} sales data.

\textbf{Autonomie:}
The simulation software \emph{Autonomie Rev 16SP7} includes detailed vehicle dynamics, energy models, and a library for several types of vehicles, which can be used for estimating fuel/energy consumption and other vehicle performances such as emissions, regenerative braking, etc.~\cite{ANL}. Each vehicle model is composed of detailed plant and controller models for its components, including engine, drivetrain, driver, and environment. The RAV4 model is based on a small SUV template, adapted to the specifications of the RAV4 \cite{ALPHA_Map}. Autonomie works with MATLAB and Simulink, where its blocks and files can be modified and customized.

\textbf{Virtual chassis dynamometer:}
In order to compute vehicle performance maps in a full parameter space of driving situations, the Autonomie model framework is modified as follows: (1)~in the driver model, braking is deactivated, and the accelerator is forced to a prescribed test pattern; (2)~in the environment model, a PI controller is inserted to artificially adjust road grade so that actual vehicle speed matches the target speed as the accelerator test pattern varies power delivered to the wheels; (3)~in the gearbox demand and transient controllers, gear shifting is deactivated; (4)~in the engine transient controller, the idle speed controller logic is bypassed to simply pass through the engine torque demand;~(5) in the wheel plant, additional load is introduced to cancel the artificial road grade signal.

The thus modified model is then run, gear-by-gear, through a complete velocity--load phase space, where the load represents various acceleration values. Each set point is held for 10 seconds to allow the model to reach local equilibrium. The resulting maps (vehicle speed to engine speed, vehicle speed and wheel force to engine torque, and engine speed and torque to fuel-rate) are then used in the next step.

\textbf{Semi-principled model:}
We formulate a simplified, semi-prin\-cipled, model that has a physics-based part (using vehicle mass, road load coefficients, gear ratios, final drive ratio, tire diameter, maximum engine torque, maximum engine speed, and engine idle speed), but it also relies on the maps obtained from the above virtual chassis dyno. The model takes as inputs the instantaneous vehicle speed $v$, acceleration $a$, and road grade $\theta$, and outputs engine speed, engine torque, fuel consumption, gear, transmission output speed, wheel force, wheel power, and feasibility of the given $(v,a,\theta)$ with respect to engine speed and engine torque. Gear scheduling is based on choosing the (feasible) gear that yields the minimal fuel consumption. In contrast to the original Autonomie model, this simplified model now yields the fuel consumption rate $f$ (and other outputs) as a direct function $f = F(v,a,\theta)$.

The model version used herein is simplifying real vehicle dynamics by assuming that the torque converter is always locked (in reality, an open torque converter bypass facilitates vehicle launch and mitigates driveline vibration). This simplification results in slight underpredictions of the fuel rate in lower gears, and it will be improved in future versions.

\textbf{Tuning parameters:}
Beyond the physics-based vehicle parameters, the simplified model also uses a few tuning parameters. These are extracted in an automated fashion from the original Autonomie vehicle model run on test cycles: minimum engine torque after gear shifting, fuel cut speed, upshifting engine speeds, and downshifting vehicle speeds.

\textbf{Fitted polynomial model:}
To generate even simpler models, a further simplification step is applied. For each vehicle, the semi-principled model is evaluated on a grid in a region of the feasible $(v,a)$-space. To these data, a (capped) degree 3 bivariate polynomial (in $v$ and $a$) is fitted (least squares sense with non-negativity constraints on the 10 parameters) of the form
\begin{align*}
f(v,a) = \max\!\Big\{&C_0 + C_1 v + C_2 v^2 + C_3 v^3 + p_0 a + \\[-.5em]
&p_1 av + p_2 av^2 + q_0 a_+^2 + q_1 a_+^2 v,\; \beta\Big\}\;,
\end{align*}
where $a_+ = \max\{a,0\}$, and $\beta$ is the minimum fuel rate, which is not necessarily zero because different vehicles have different criteria for enacting a fuel cut.

\begin{figure}[htbp]
\includegraphics[width=\linewidth]{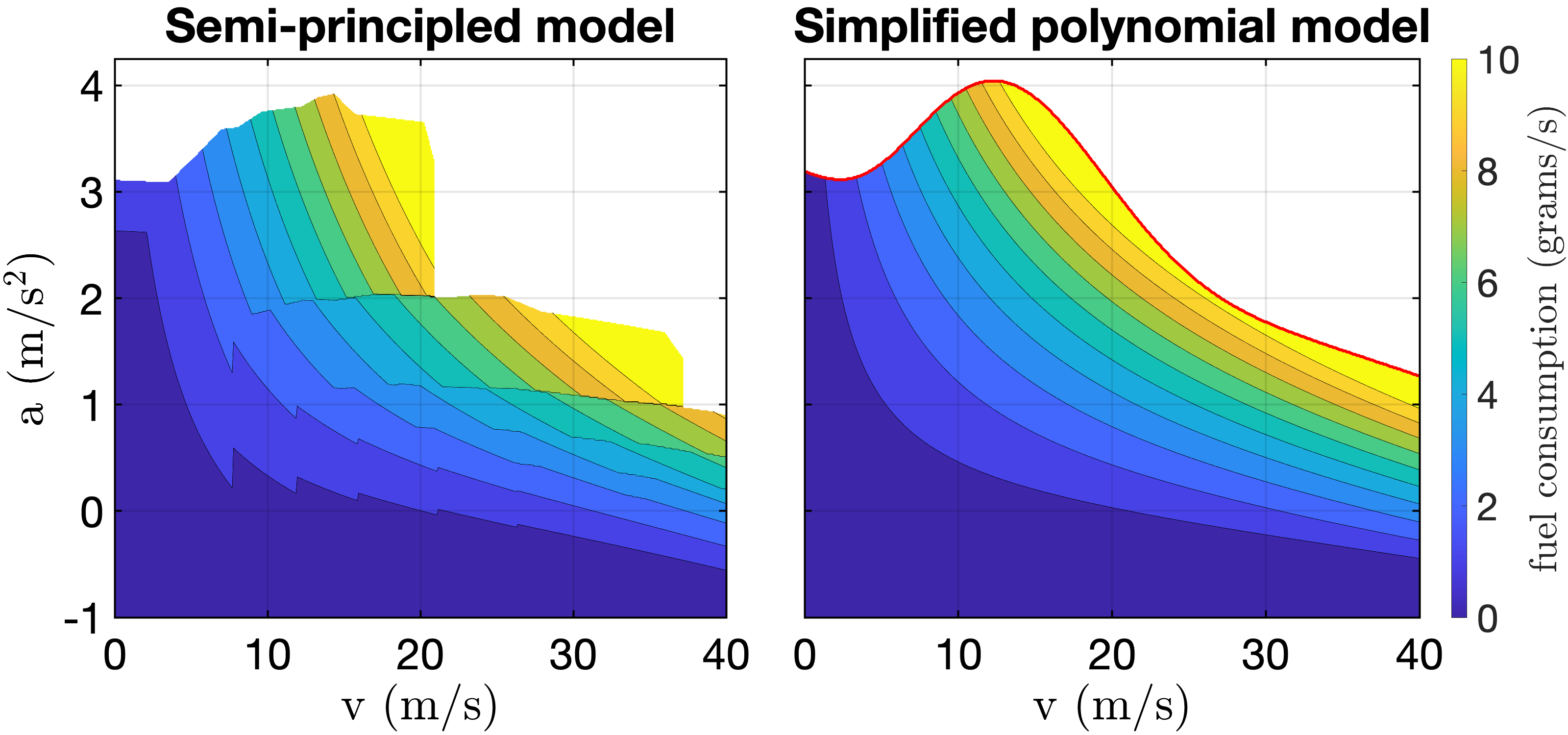}
\vspace{-2em}
\caption{Semi-principled fuel consumption model (left) and simplified polynomial model (right) for midsize SUV. The red curve is \boldmath$g(v)$, the boundary of the feasiblity region.}
\label{fig:midSUV-models}
\end{figure}

Moreover, a fit for the boundary of the feasibility region is produced, in the form of a function $g(v)$, above which $(v,a)$-pairs are infeasible. 
Figure \ref{fig:midSUV-models} shows a plot of the semi-principled and fitted polynomial models for a midsize SUV.

\textbf{Model validation:}
For each vehicle, the accuracy of both model simplifications is validated against the Autonomie ground truth based on standard drive cycles (for that vehicle type). Used are: (a)~the EPA cycles UDDS, FTP-75, US06, 505 for light-duty vehicles \cite{EPA_cycles}; (b)~the globally (UNECE) approved WLTC \cite{WLTC}. Table~\ref{table:model_list_and_errors} summarizes, for each vehicle, the relative error of both model types relative to Autonomie, averaged over all respective drive cycles. One can see that on average, all simplified models are safely within 4\% relative error.

\begin{table}
\begin{center}
\begin{tabular}{C{3.4cm} C{1.2cm} C{1.2cm} C{1.0cm} }
\toprule
vehicle model & model & model & HV \\
(engine type) & error SP & error FP & share \\
\midrule
Compact Sedan (SI) & $-2.0\%$ & $-2.7\%$ & $23.59\%$ \\ 
Midsize Sedan (SI) & $\phantom{-}2.7\%$ & $\phantom{-}2.9\%$ & $32.92\%$ \\ 
Midsize SUV (SI) & $-2.2\%$ & $-2.3\%$ & $17.56\%$ \\ 
Midsize Pickup (SI) & $-3.1\%$ & $-3.0\%$ & $10.32\%$ \\ 
Class3 PND (CI) & $\phantom{-}0.2\%$ & $-2.1\%$ & $15.61\%$ \\ 
\midrule
2019 RAV4 (SI) & $-1.6\%$ & $-1.0\%$ & -- \\ 
\bottomrule
\end{tabular}
\end{center}
\caption{Vehicle models. Error SP (FP) is the relative error of the semi-principled (fitted polynomial) model vs.\ Autonomie baseline.
(HV=human vehicle; SI=spark-ignition; CI=compression-ignition; PND=pickup-and-delivery)}
\label{table:model_list_and_errors}
\end{table}

\section{Integration and Simulation Results}
\label{sec:integration}
The major components of microsimulation that SUMO handles are \textit{car-following}, \textit{lane-changing}, and \textit{routing}. We have built an interface layer on top of SUMO using Flow~\cite{flow}, which serves as the backbone infrastructure of our integrated framework. Flow builds on SUMO's underlying logic to allow both for custom definition of car-following logic, and for the implementation of vehicle level control schemes, such as in the form of \emph{reinforcement learning} (RL). Through Flow, microsimulations with custom network geometries are instantiated in SUMO and interfaced with different vehicle controllers and behavior models, road networks, traffic environments, vehicle energy models, etc. This framework is modular in its components and features, thereby enabling a systematic assessment of different choices of model design.

Simulation outputs are post-processed in a custom data pipeline implemented in Amazon AWS cloud. Output data are uploaded to AWS S3 storage as CSV files, which then triggers automatic post-processing using AWS Lambda and AWS Athena services. A load-balancing mechanism is implemented to handle parallel uploads at any scale. Centralized storage allows for standardized post-processing that is decoupled from simulating vehicle dynamics; \emph{i.e.}, simulations do not need to be rerun as post-processing procedures and models are updated. SQL queries are used to compute several \emph{key performance indicators} (KPIs). Resultant table schemata are intentionally designed to facilitate quick file ingestion and plot rendering by an online benchmarking leaderboard for ranking controller strategies. The leaderboard, built with DataTables \& \texttt{Plotly.js}, features standardized scenarios, evaluation metrics and plots, and vehicle distributions.

The primary KPI of this study is fuel economy (\emph{i.e.}, \emph{miles per gallon} (mpg)) of the whole traffic flow, leveraging the energy models from Section \ref{sec:energy} to estimate fuel consumption\footnote{The energy models of Section \ref{sec:energy} are implemented twice in: (1) Flow for RL optimization; (2) AWS for post-processing evaluation.}. Further, the evaluation pipeline has implemented several vehicle distributions by mapping vehicles to energy models. CAVs are mapped to the RAV4 model, and others are mapped onto the ``human'' vehicles according to proportions shown in Table~\ref{table:model_list_and_errors}. Several other KPIs (\emph{e.g.}, network speed, network inflow rate) and plots (\emph{e.g.}, time-space diagrams, mpg histograms, various telemetry vs.\ relative time or distance) further illustrate each controller's performance relative to a standardized baseline of uncontrolled traffic flow with waves.

\textbf{Experimental setup:}
The benchmarking pipeline is designed for a broad customizable range of scenarios, parametric studies, and vehicle distributions. The scenario discussed herein features a network model of CA SR-134 (\textasciitilde 1-mile stretch from the I-210 Pilot project \cite{ConnectedCorridors}). The road is assumed flat, and ramps are disengaged to isolate the effects of waves due to congestion. All human-driven vehicles are dynamically equivalent, utilizing the IDM parameters from Section \ref{sec:waves} and SUMO's lane-changing. User-defined, longitudinally-controlled CAVs are evenly injected at prescribed penetration rates. Edges are added upstream and downstream to remove boundary effects from metrics evaluation. Unless unsafe, vehicles are injected at $2050$ vehicles per hour (a value that was found to lead to both reasonable congestion and desirable wave formation). Congestion is achieved by imposing a $5$ m/s speed limit immediately downstream from the simulation domain. The simulator uses a ballistic integration scheme with a time step of $0.4$ s. Zero-mean Gaussian noise of standard deviation $0.1$ m/s\textsuperscript{2} is added to the acceleration to provoke waves, as previously discussed. The simulation is warmed up for 720 s to let waves establish throughout the domain and then run for 1200 s.

\textbf{Sparse Lagrangian traffic controllers:}
As a demonstration of the framework, two parameterized vehicle controllers are submitted through the pipeline at various penetration rates ($5\%$--$10\%$): (1)~FollowerStopper (FS)~\cite{stern2017dissipation} with varied $v_\text{desired}$ ($3$ m/s--$7$ m/s) 
and (2)~IDM with Relaxation (IDM+R)~\cite{cui2017stabilizing}, given by $\dot{v} = f_\text{IDM}(s, v, \Delta v) + \gamma (v_\text{desired}-v)$. For simplicity, the human CFM parameters in \eqref{eq:IDM} are assumed precisely known here. The control gain $\gamma$ is a tunable parameter. Here we study varying $v_\text{desired}$ ($3$ m/s--$7$ m/s) and $\gamma$ (0.5 /s \& 1.0 /s).

\begin{figure}
\includegraphics[width=\linewidth]{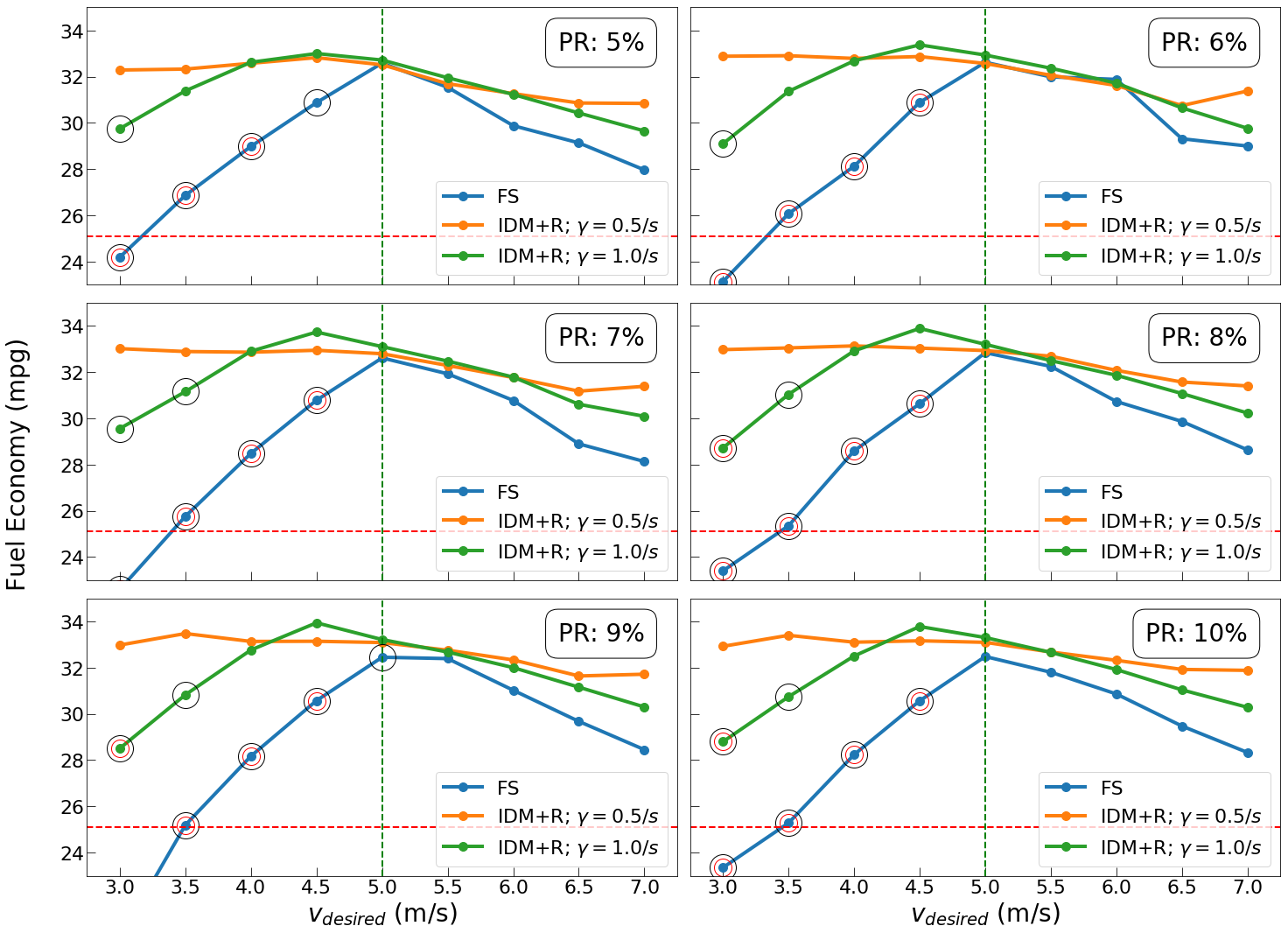}
\vspace{-2em}
\caption{Fuel economy results of three vehicle controllers over parameter \boldmath$v_\text{desired}$ (x-axis) and penetration rates (subplots). Red (black) circles denote decreased inflow rate (network speed) by more than 10\%. Red dashed line denotes baseline fuel economy, and green dashed line highlights the true average traffic speed downstream.}
\label{fig:controller_results}
\end{figure}

\textbf{Results:}
As shown in Figure \ref{fig:controller_results}, as long as $v_\text{desired}$ is a reasonable estimate of the true average speed of traffic, all tested controllers result in significant energy improvements, with the exception of the FS for $v_\text{desired} < 3.5$ m/s. As the flagging thresholds in Figure~\ref{fig:controller_results} show, the FS sacrifices on network speed and throughput when $v_\text{desired}$ is too low, while the IDM+R controller generally appears more robust in those regards, yet still performs well in fuel economy. Furthermore, the IDM+R controller consistently matches or outperforms the FS. The IDM+R controller has a free parameter $\gamma$ (how aggressively to steer $v$ towards $v_\text{desired}$). The results suggest that more aggressive control (\emph{i.e.}, larger $\gamma$) has equal or positive impact if $v_\text{desired}$ does not significantly underestimate the average speed. Otherwise, the more aggressive controller suffers from similar drawbacks as the FS. 
Conversely, overestimating the true average speed results in less efficient wave dampening.
\begin{figure}
\includegraphics[width=\linewidth]{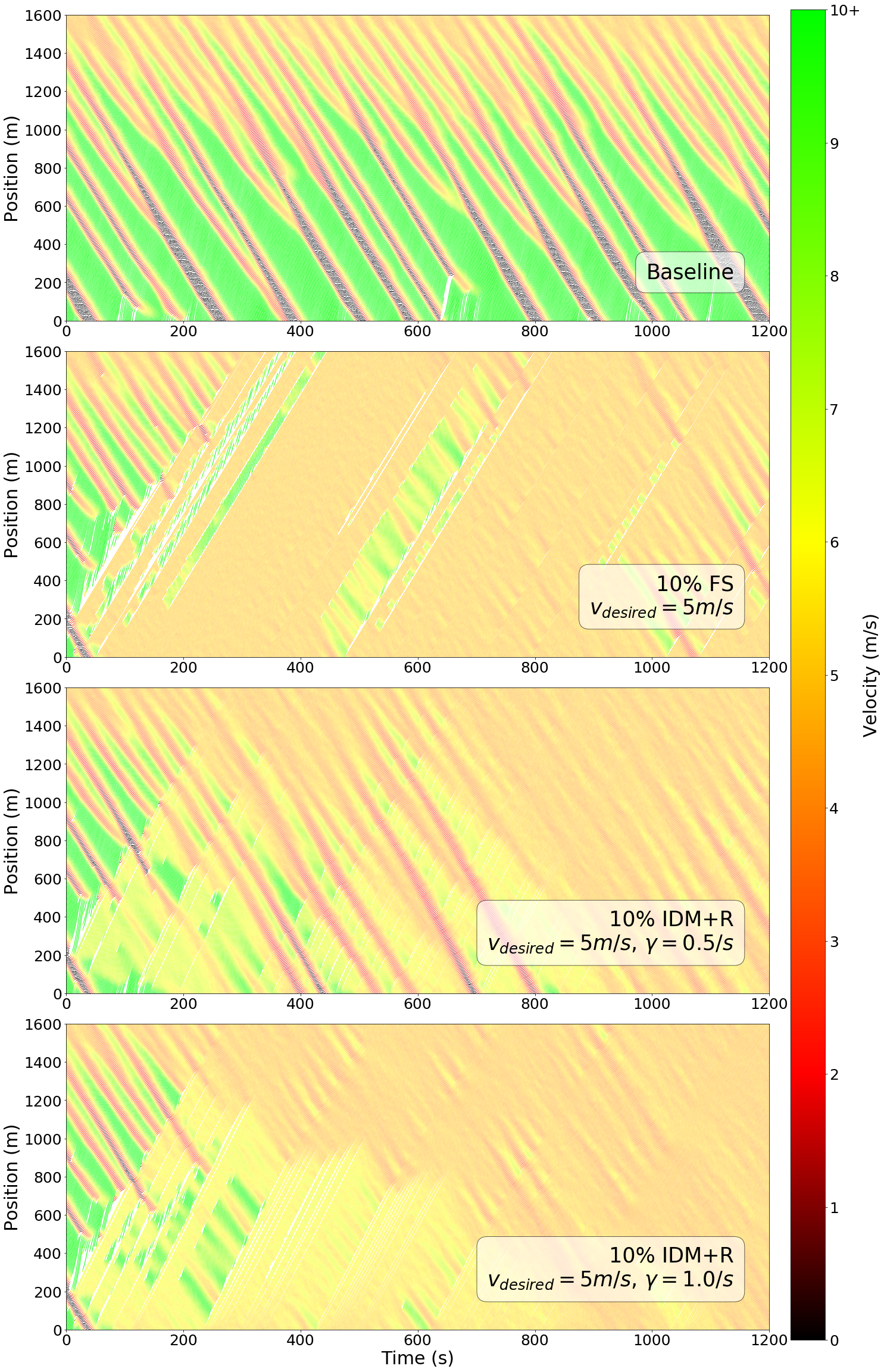}
\vspace{-2em}
\caption{Time-space diagrams of controllers at 10\% penetration rate and \boldmath$v_\text{desired}=$ 5 m/s, which yield similar fuel economy. The uncontrolled baseline shows unimpeded waves; controlled cases show some wave dissipation.}
\label{fig:controller_tsd2}
\end{figure}

\begin{figure}
\includegraphics[width=\linewidth]{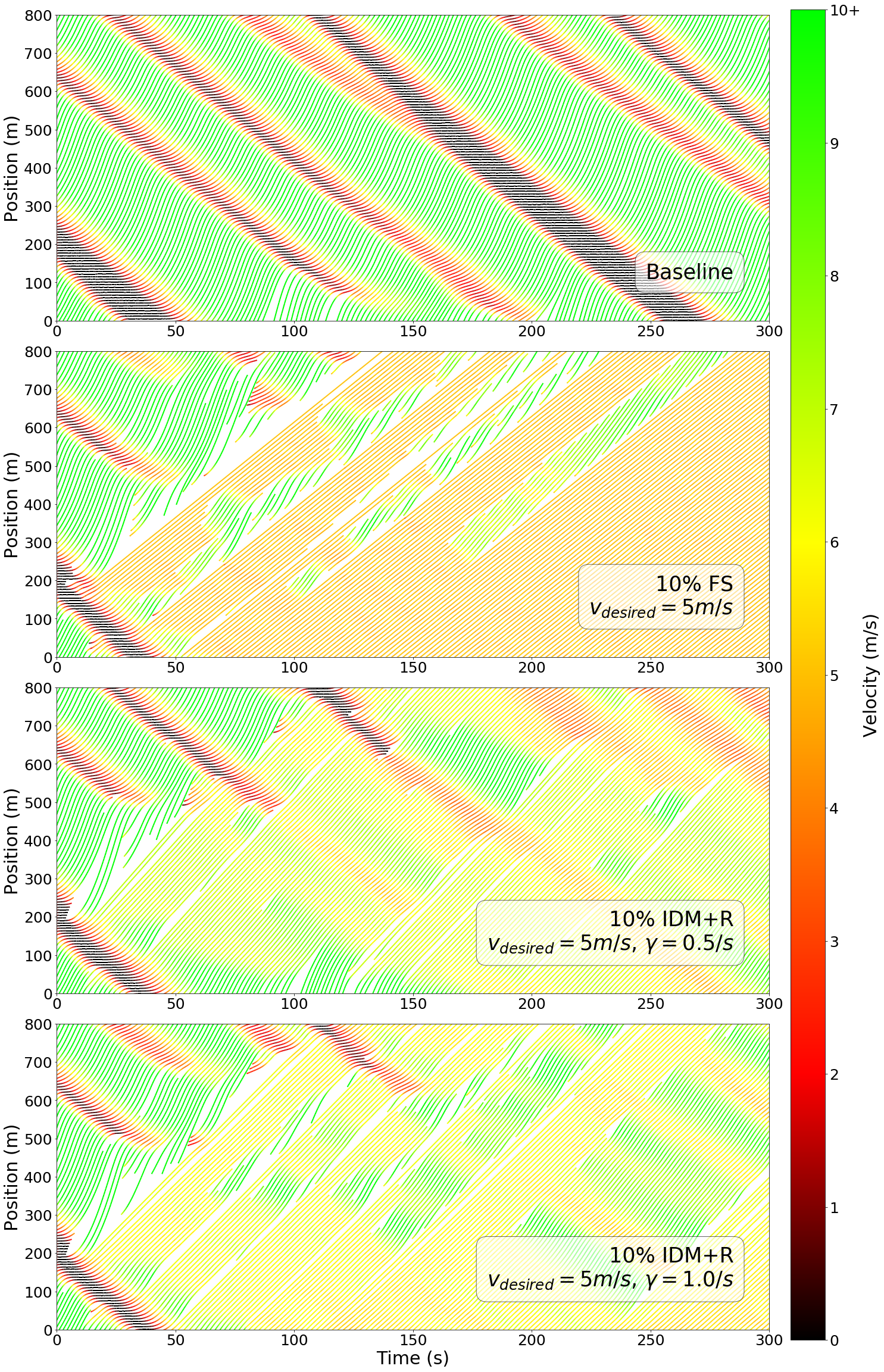}
\vspace{-2em}
\caption{The same time-space diagrams in Figure \ref{fig:controller_tsd2} zoomed in on both axes to highlight individual trajectories.}
\label{fig:controller_tsd}
\end{figure}
Figure \ref{fig:controller_tsd2} \& \ref{fig:controller_tsd} shows how the controllers act to interrupt the progression of
waves via their increased tendencies to drive at uniform speeds. It is
apparent that no controller removes all waves, and that there are
qualitative differences across the different controllers. Most
prominently, the FS conducts noticeably more wave removal than the
IDM+R controllers; however, it does so at the expense
of allowing more lane changes. These aspects result in an overall
similar performance in terms of energy metrics.

\section{Conclusions and Outlook}
\label{sec:conclusions}
The presented framework provides a standardized benchmarking tool for researchers to develop and assess Lagrangian traffic-smoothing controllers. Neither of the presented controllers were constructed to be optimal; rather they are simple recipes taken from existing work that prove the concept. The authors present this work as a vision of how future optimized controllers may be evaluated against one another.


\begin{acks}
The authors would like to thank Kenneth Butts (Toyota) for helpful discussions and comments. This material is based upon work supported by the National Science Foundation under Grants CNS-1837244/CNS-1837652/CNS-1837481/OISE-1743772. This material is based upon work supported by the U.S.\ Department of Energy’s Office of Energy Efficiency and Renewable Energy (EERE) under the Vehicle Technologies Office award number CID DE--EE0008872. The views expressed herein do not necessarily represent the views of the U.S.\ Department of Energy or the United States Government.
\end{acks}

\bibliographystyle{ACM-Reference-Format}
\bibliography{circles_refs}

\end{document}